\documentclass[a4paper]{article}

\usepackage{icrc2013}
\usepackage[english]{babel}

\title{Upgrade of the MAGIC telescopes}

\shorttitle{Upgrade of the MAGIC telescopes}

\authors{
Daniel Mazin$^{1,2}$,
Diego Tescaro$^{2,3,5}$,
Markus Garczarczyk$^{4}$,
Gianluca Giavitto$^{2,6}$,
Julian Sitarek$^{2}$
for the MAGIC Collaboration.
}

\afiliations{
$^1$ Max-Planck-Institut f\"ur Physik, D-80805 M\"unchen, Germany \\
$^2$ IFAE, Edifici Cn., Campus UAB, E-08193 Bellaterra, Spain \\
$^3$ INFN Pisa, I-56127 Pisa, Italy \\
$^4$ Deutsches Elektronen-Synchrotron (DESY), D-15738 Zeuthen, Germany \\
\scriptsize{
$^{5}$ now at: Inst. de Astrof\'{\i}sica de Canarias, E-38200 La Laguna, 
and Universidad de La Laguna (ULL), Dept. Astrof\'{\i}sica, E-38206 La Laguna, Tenerife, Spain \\
$^{6}$ now at: Deutsches Elektronen-Synchrotron (DESY), D-15738 Zeuthen, Germany.
}
}

\email{mazin@mpp.mpg.de}

\abstract{
The MAGIC telescopes are two Imaging Atmospheric Cherenkov Telescopes (IACTs)
located on the Canary island of La Palma. With 17m diameter mirror dishes and
ultra-fast electronics, they provide an energy threshold as low as 50 GeV for
observations at low zenith angles. The first MAGIC telescope was taken in
operation in 2004 whereas the second one joined in 2009. In 2011 we started a
major upgrade program to improve and to unify the stereoscopic system of the
two similar but at that time different telescopes. Here we report on the
upgrade of the readout electronics and digital trigger of the two telescopes,
the upgrade of the camera of the MAGIC~I telescope as well as the commissioning
of the system after this major upgrade.}

\begin{document}
\maketitle

%Begin a section.
\section{Introduction}

\begin{figure}
    \includegraphics[width=0.49\textwidth]{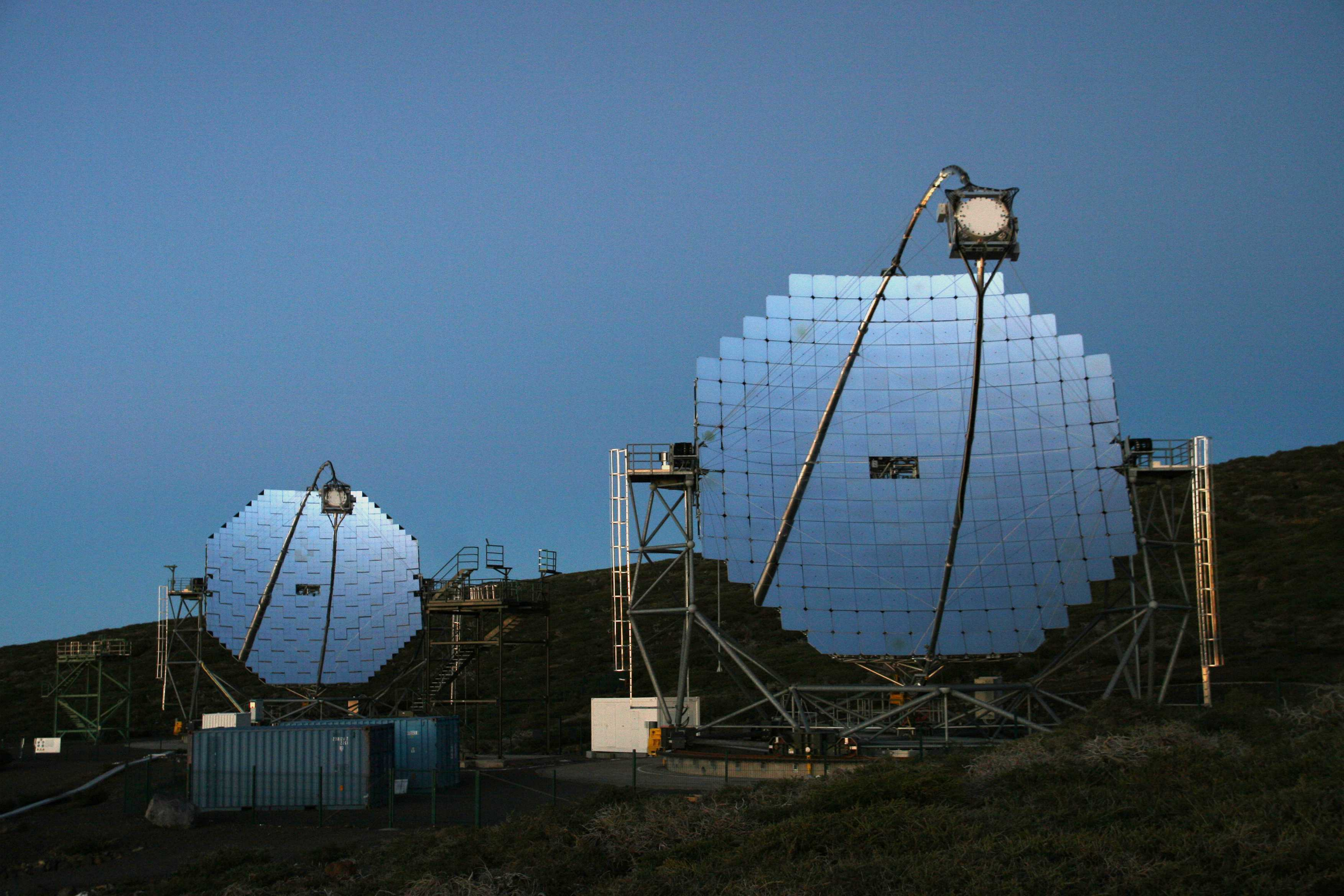}
    \caption{The two 17m diameter MAGIC telescope system operating at the
Roque de los Muchachos observatory in La Palma.
The front telescope is the MAGIC~II.}
    \label{2-Magics-photo}
\end{figure}
\setcounter{footnote}{0}   
 
The MAGIC telescopes (see Fig.~\ref{2-Magics-photo}) is a stereoscopic system of two Imaging 
Atmospheric Cherenkov Telescopes (IACTs) located at the observatory
of Roque de los Muchachos in La Palma, Canary Islands. 
The telescopes are designed to observe very high energy gamma rays
at energies between 50 GeV and tens of TeVs.
The two MAGIC telescopes have been constructed 5 years apart (2004 and 2009, respectively),
which, due to the technological progress along the time, led to rather significant differences in the two systems.
 \begin{itemize}
   \item The camera of the MAGIC~I telescope consisted of 577 pixels (divided in small pixels, 1 inch diameter,
in the inner part of the camera and large pixels, 2 inch diameter, in the outer part).
The camera of MAGIC~II consists of 1039 pixels, all small, 1 inch diameter.
   \item The trigger area of MAGIC~II had an affective area of 1.66 times larger than the one of MAGIC~I.
   \item  MAGIC~I readout was based on
optical multiplexer and off-the-shelf FADCs (MUX-FADC \cite{muxfadc}), which was robust and had an excellent performance
but was expensive and bulky). The readout of MAGIC~II was based on the DRS2 chip \footnote{see http://drs.web.psi.ch/}
(compact and inexpensive but performing worse in terms of intrinsic noise, dead time and linearity compared to the MUX-FADC system).
   \item Receiver boards of MAGIC~I, which are responsible for converting the optical signals into electrical ones, to split the signal into the analogue
branch of the readout and the digital branch for the trigger, as well as to apply a Level-0 trigger condition (L0-trigger)\footnote{The Level-0 trigger is a discriminator
applied on a signal amplitude.}
depending on the height of the signal in a trigger channel, were built using an old technology and were showing high failure rate (due to aging).
This resulted in frequent problems in setting the discriminator thresholds (DTs) during the observations and only a rather slow reaction
of the individual pixel rate control (IPRC), in order of several minutes, to adjust the DTs depending on the star field of the observations. This caused some loss of
costly observation time. The receiver boards of MAGIC~II allow instead a fast response (order of several seconds) of the IPRC 
thanks to a build-in FPGA and an easy and robust communication through the VME bus.
 \end{itemize}
In years 2011-2012 MAGIC underwent a major upgrade program to
improve and to unify the stereoscopic system of the two telescopes. This paper describes the main items of the upgrade
and the commissioning of the upgraded MAGIC.

\section{Individual steps of the upgrade}

%-------------------------------------------
\begin{figure}[phtb]
\begin{center}
\includegraphics[width=1.0\linewidth]{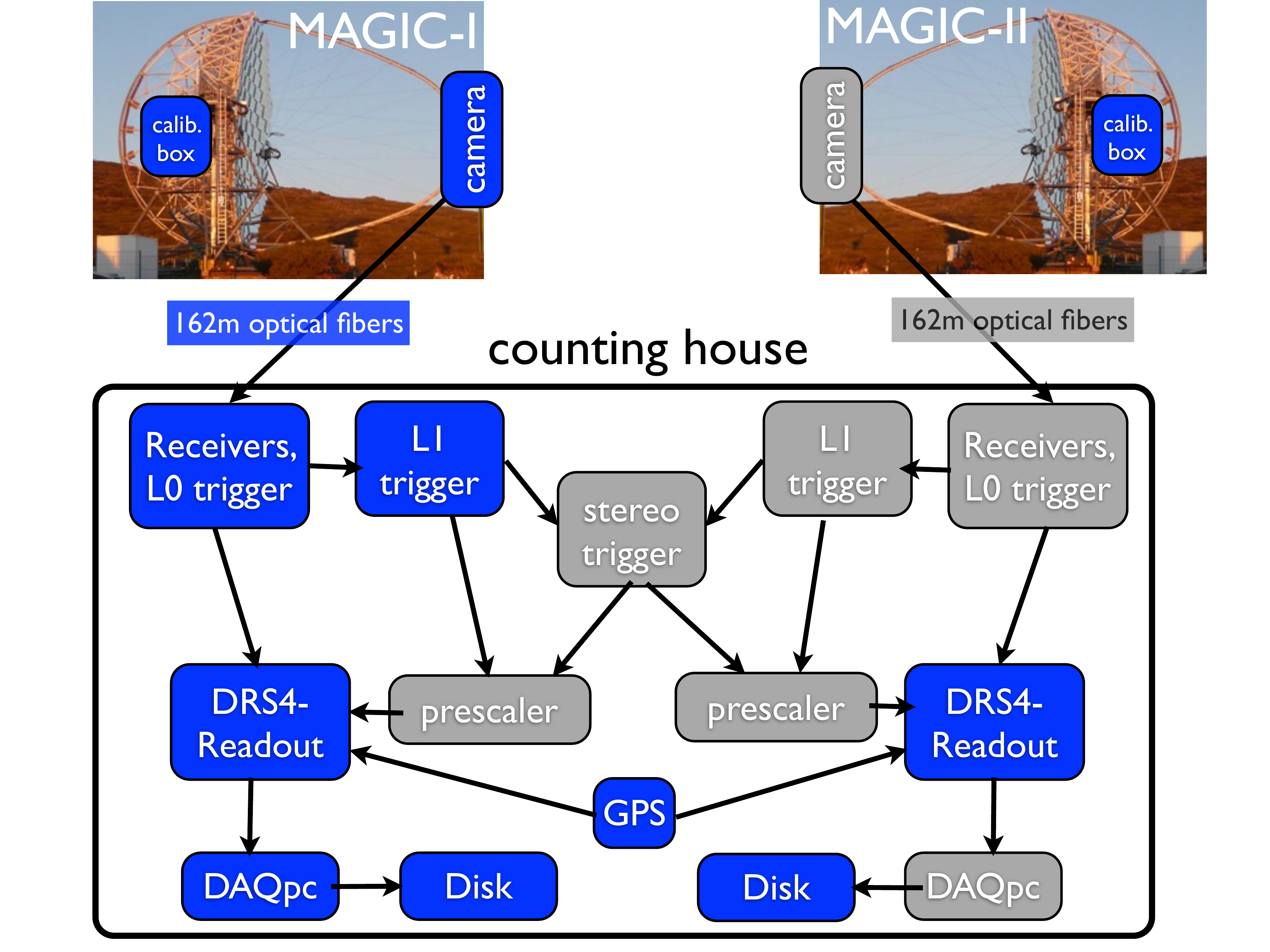}
\caption{
Schematic view of the readout and trigger chain of the MAGIC telescopes.
The blocks in the blue boxes have been replaced and commissioned during the upgrade.}
\label{fig:MAGIC-Scheme}
\end{center}
\end{figure}
%-------------------------------------------

The main goals of the upgrade were replacing the camera and the Level-1 trigger (hereafter L1, coincidence trigger for a group of neighboring pixels) of the MAGIC~I telescope
%to be the same as of the MAGIC~II telescope
as well as changing the readout of the two telescopes. %unifying the readout of the two telescopes.
The readout was changed to a system based on the
DRS4 chip, sampling the signals with 2 Gsamples/s.  The new readout is cost effective, has a linear behavior over a
large dynamic range, less than 1\% dead time and low noise and negligible chgannel-to-channel cross-talk (\cite{bitossi,icrc_performance}). 
This allowed us to maintain the quality of the
readout based on MUX-FADCs while reducing cost and saving space as we needed to accommodate the readout for more than 2000 channels
(the two MAGIC telescopes) into the space available in the control house of the experiment where the readout and the trigger are located.
The receiver boards and the trigger of MAGIC~I have also been upgraded to be equal to the ones of MAGIC~II.
The individual items of the upgrade program are shown in Fig.~\ref{fig:MAGIC-Scheme}. 
The blue boxes represent hardware items which have been replaced and commissioned during the upgrade.

%-------------------------------------------
\begin{figure}[phtb]
\begin{center}
\includegraphics[width=0.8\linewidth]{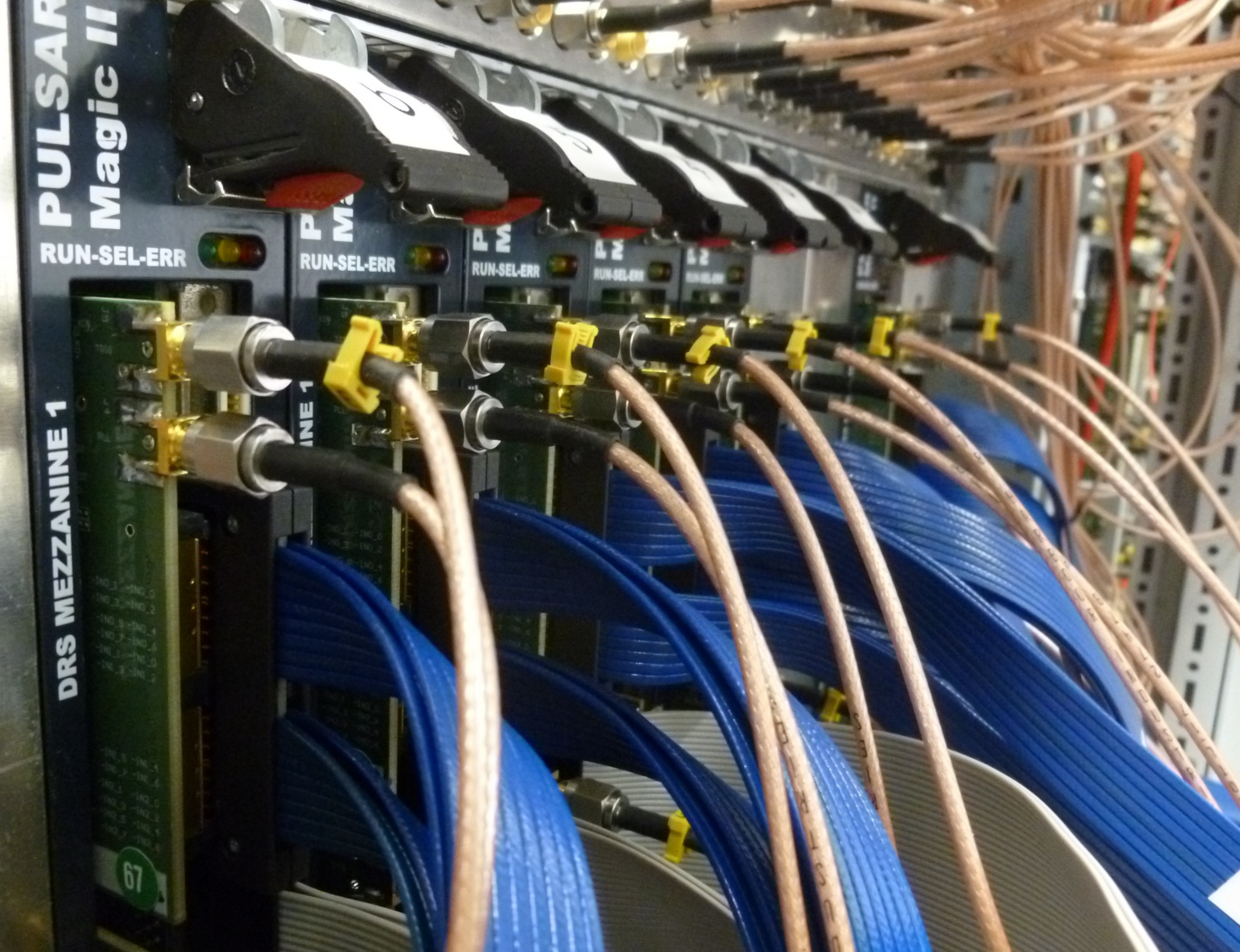}
\caption{
A close-up into the electronics, which read out the
ultrafast signals produced by the Cherenkov photons from cosmic-ray induced atmospheric showers in the camera of the
MAGIC telescopes.}
\label{fig:MAGIC-Readout}
\end{center}
\end{figure}
%-------------------------------------------

The upgrade was staged. In the first stage, which required a
shutdown of the telescopes between June and November 2011,
the following tasks were performed:
 \begin{itemize}
  \item Restructuring of the electronics and the computing rooms. 
%Whereas the trigger and the readout electronics
%shared the same room before and things became crowded with years,
A dedicated computing room was created for the subsystems machines
as well as for the analysis farm. The electronics room was instead completely renewed,
including installing new electrical lines, a new cooling system,
closed racks for electronics and second floor for better cable routing.
  \item The readout electronics of the MAGIC~I telescope was changed from
MUX-FADC to a system based on the DRS4 chip (see Fig.~\ref{fig:MAGIC-Readout}). % , which is much more compact and flexible.
  \item The readout electronics of the MAGIC~II telescope was changed from
the one based on DRS2 chip to a system based on the DRS4 chip (see for the advantages above and in \cite{sitarek}).
As a major goal it was important to unify the readouts of the two telescopes to ease maintenance
and increase the robustness.
  \item The calibration system of the two telescopes were upgraded.
  \item Upgrade of the computing system, which was largely renewed
to increase the computing power, disk storage capacity and maintainability.
 \end{itemize}
The MAGIC telescopes restarted physics programm by December 2011, which lasted until June 2012 when the
second part of the upgrade took place.

%-------------------------------------------
\begin{figure}[phtb]
\begin{center}
\includegraphics[width=1.0\linewidth]{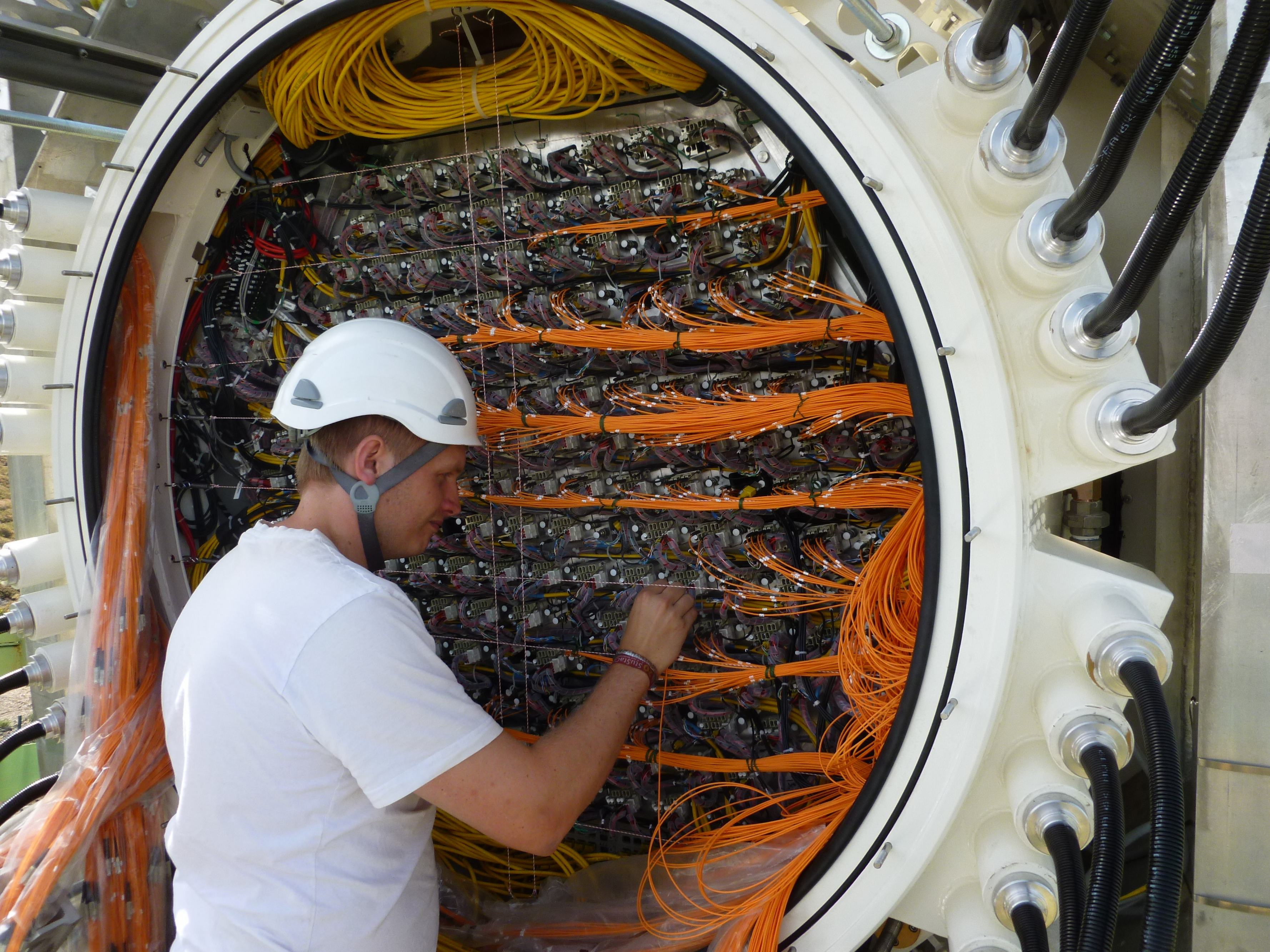}
\caption{
Cabling of the new camera at the MAGIC~I
telescope. Electrical signals at the camera sensors are
converted into optical ones and transmitted over
optical fibers %(oranges cables in the picture) 
to the readout electronics at the control room.}
\label{fig:MAGIC-CamInstallation}
\end{center}
\end{figure}
%-------------------------------------------

In the second phase of the upgrade, which resulted in a scheduled shutdown between mid June 2012
and end of October 2012, the following tasks were achieved:
  \begin{itemize}
   \item Exchange of the camera of the MAGIC~I telescope (see Fig.~\ref{fig:MAGIC-CamInstallation})
by a clone camera of the MAGIC~II telescope.
This also meant an exchange of the optical fibers and power cables of the MAGIC~I telescope as
the number of channels increased from 577 to 1039. A rearrangement of the counter weights in the structure
took also place to compensate the weight increase of the camera.
   \item Upgrade of the readout of the MAGIC~I telescope from 577 to 1039 channels\footnote{In the current configuration the readout can host up to 1152 channels per telescope}.
   \item Upgrade of the L1 trigger of the MAGIC~I telescope. The new trigger is a clone
of the MAGIC~II L1 trigger and has a 66\% larger area than the previous trigger.
The increase of the trigger area provides a more homogeneous gamma-ray
efficiency across the camera, which is especially important for
observations of extended sources and off-center observations.
 \end{itemize}

%%-------------------------------------------
\begin{figure}[htb]
\begin{center}
\includegraphics[width=0.49\linewidth]{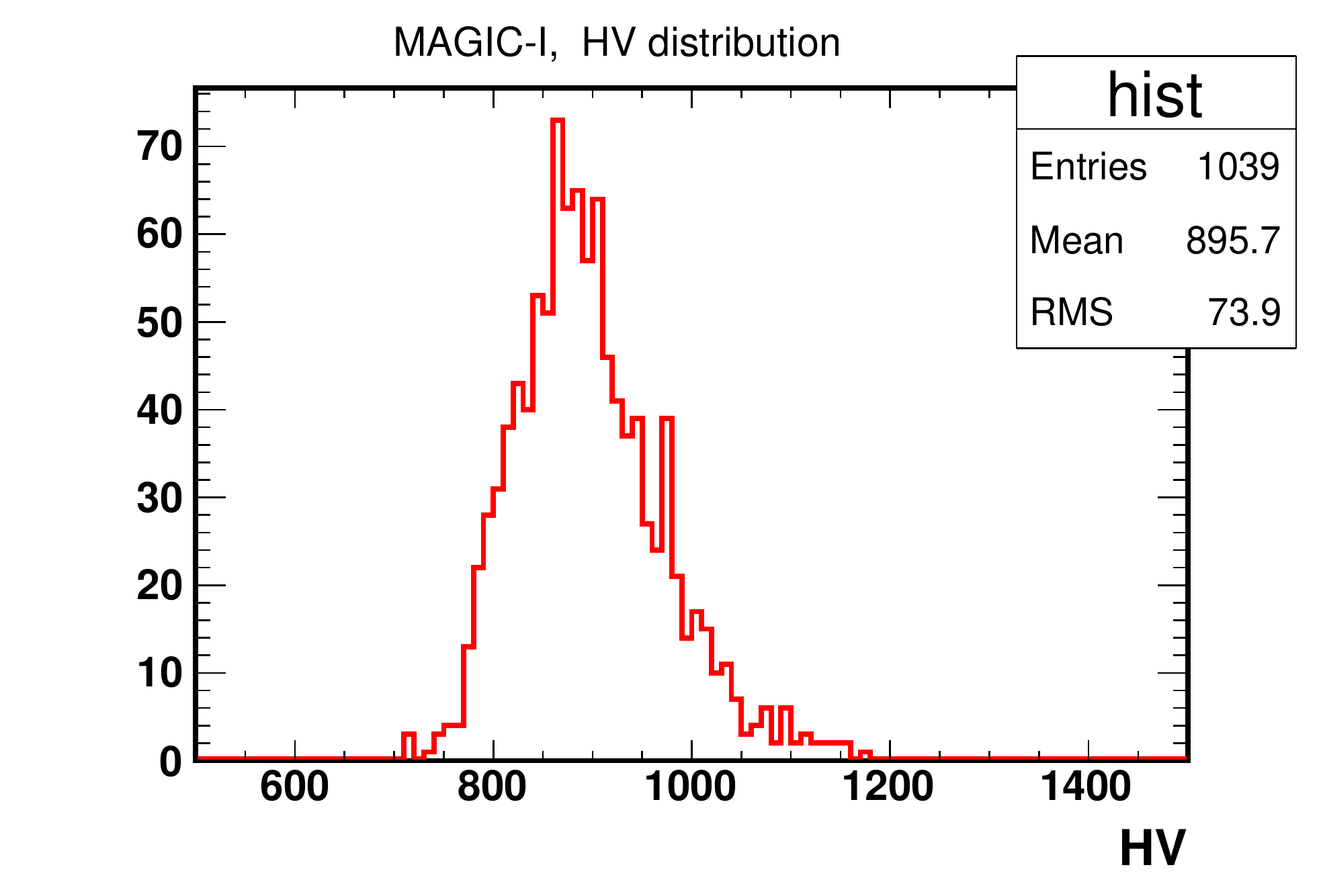}
\includegraphics[width=0.49\linewidth]{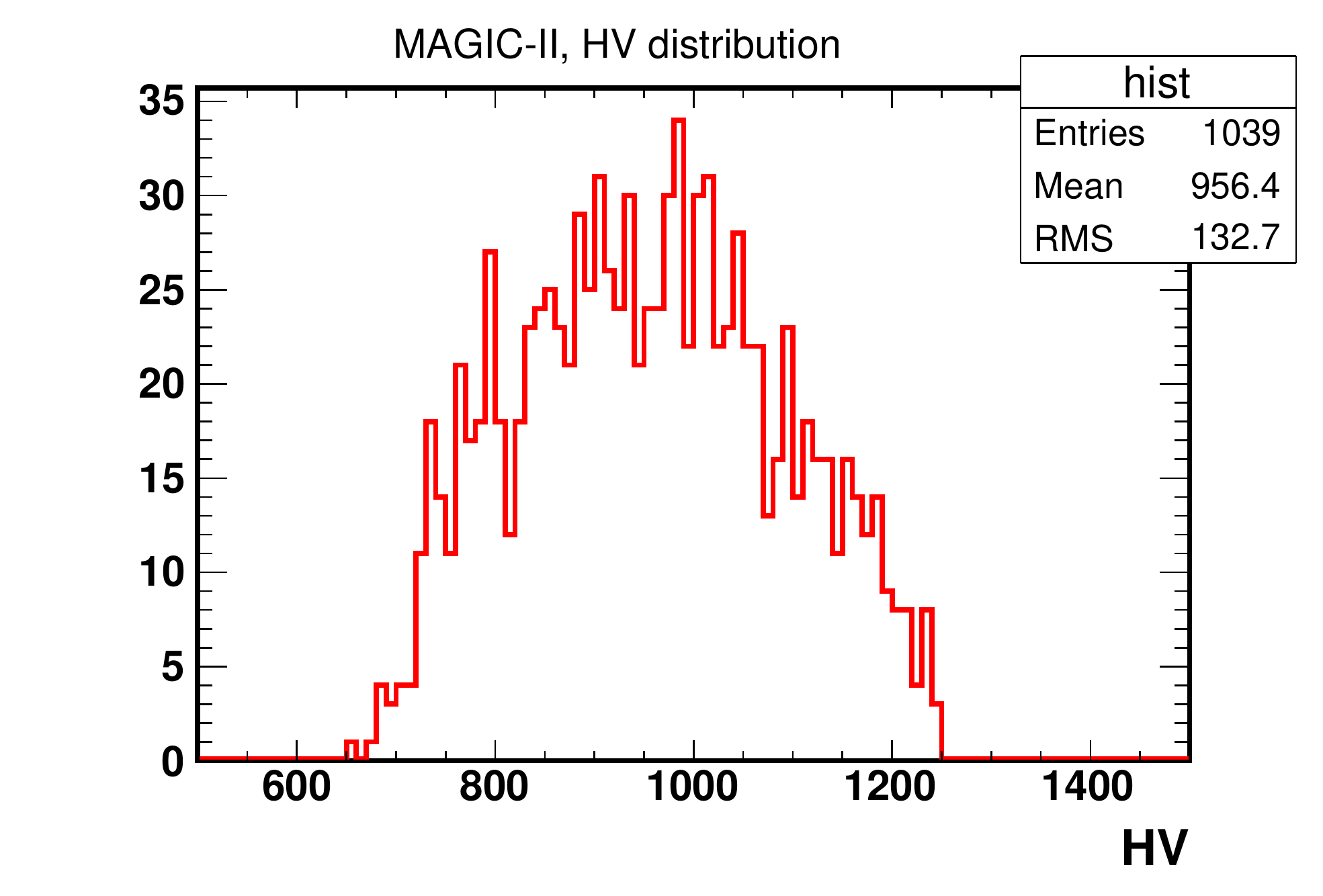}
\caption{
Distribution of the high voltages (HVs) applied to PMTs in MAGIC~I and MAGIC~II cameras after the charge flatfielding procedure.
One can see that the HV distribution in the MAGIC~II camera is wider.
Maximal voltage which can be applied to the MAGIC PMTs is 1250\,V.}
\label{fig:HVs}
\end{center}
\end{figure}
  
In addition, two resting pillars were installed for the MAGIC~I and MAGIC~II cameras in
Summer 2012. The cameras of the telescopes rest on top of the corresponding pillar while in
the parking position, which greatly improves the stability of the telescope
structure during storms, snowfall and strong winds.

\section{Commissioning of the system}

\begin{figure}[htb]
\begin{center}
\includegraphics[width=1.0\linewidth]{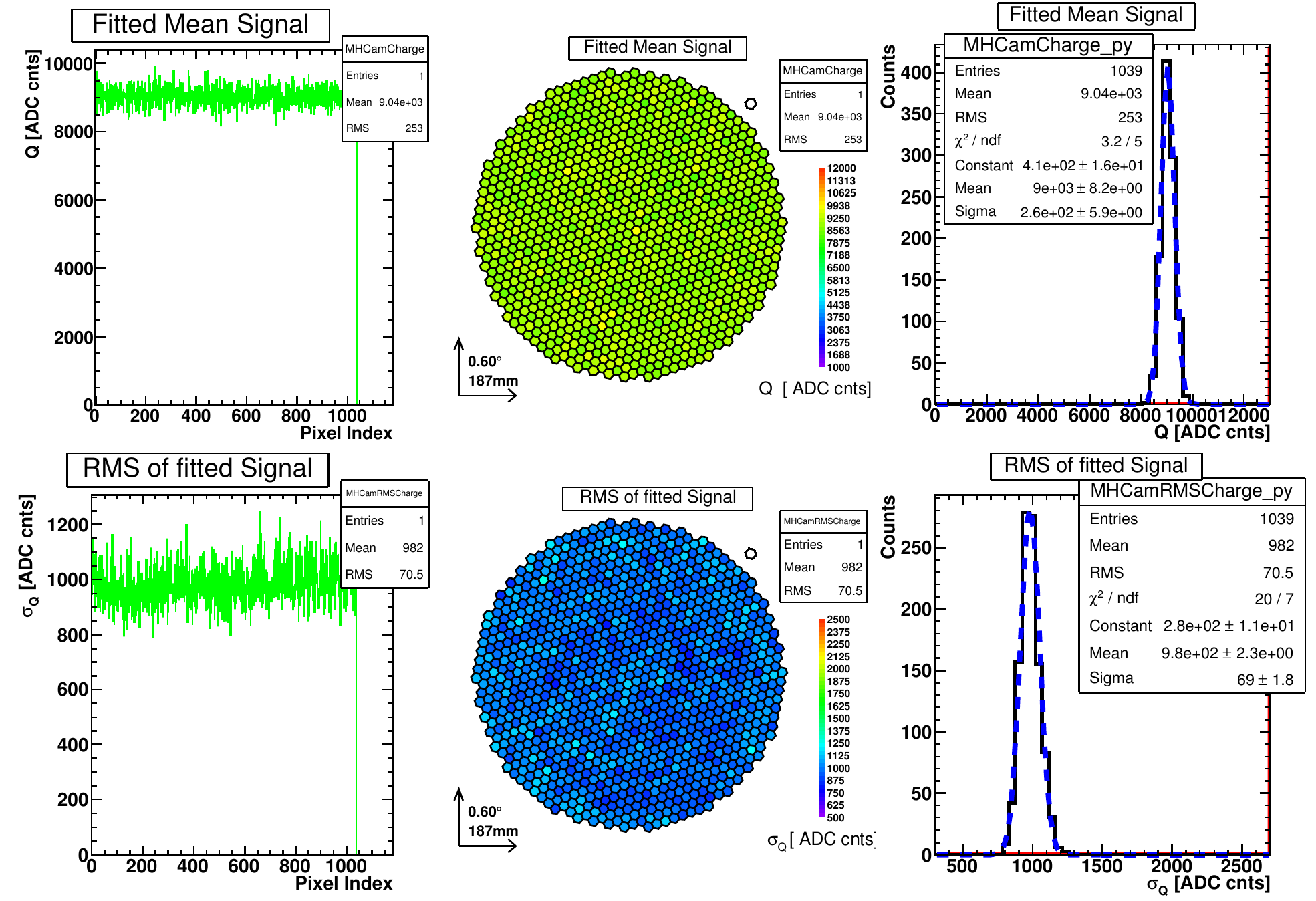}
\includegraphics[width=1.0\linewidth]{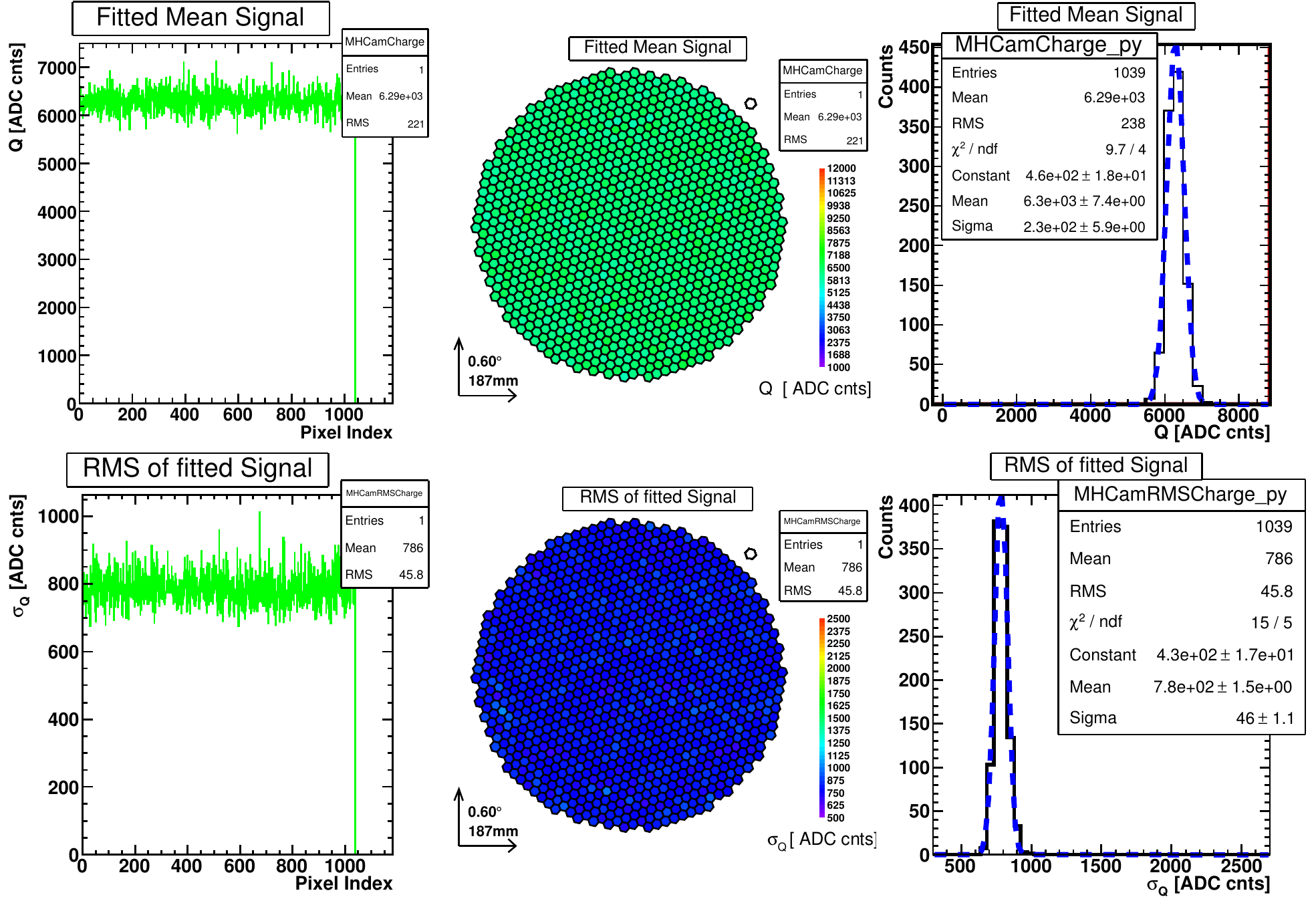}
\caption{
Charge distribution (mean and RMS) in MAGIC~I (top) and MAGIC~II (bottom) cameras after HV flatfielding for 2000 calibration pulses.
Data from 22 October, 2012.
}
\label{fig:charge}
\end{center}
\end{figure}
%-------------------------------------------
The commissioning of the upgraded system required 
a dedicated, well experienced and highly motivated team
of 5 to 10 physicists to stay on La Palma at the site of the experiment  
for a duration of several months after the installation of the hardware.
The work has been mainly devoted to the following items: \\
{\bf Mapping of the channels.} The 1039 pixels of the PMT camera have to be connected individually
to the corresponding optical fibers at the camera side. The other ends of the fibers are
then connected individually to the receiver channels. It is unavoidable that some mis-mappings
take place. In order to correct mistakes, identify broken channels and other problems in the signal transmission, 
a dedicated pulse injection system of the camera is used. 
Artificial test pulses are generated at the base of the PMTs so that the whole 
signal transmission chain, including the trigger and readout, can be tested during the day.
\\
{\bf L1 trigger check.} The L1 trigger consists of 19 macrocells, each has 36 channels. Following logic algorithms are implemented:
2 next neighbor trigger (2NN), 3NN, 4NN, and 5NN. The L1 trigger is then
programmed as an OR of the macrocells trigger.
A dedicated hardware and software have been built to test all multiplicities. The
L1 trigger systems of both telescopes have been extensively tested, hardware mistakes identified and repaired.
\\
{\bf HV flatfielding.} Each PMT has a different gain at a fixed HV. 
The spread of the gains is unavoidable during the manufacturing process and is around 30-50\%. 
The signal propagation chain introduces further differences
in the gain: the optical links as well as the PIN diodes of the receivers mainly contribute to them.
For the purpose of easier calibration of the signals, the HVs applied to PMTs are adjusted such that the resulting signal from calibration pulses 
(equal photon density at the entrance of the PMTs) is equal in all pixels when extracted after the digitization process. 
The resulting HV distribution for MAGIC~I and MAGIC~II cameras can be seen in Fig.~\ref{fig:HVs}. 
The distribution of the MAGIC~I camera is narrower. This is due to the fact that during the construction of the MAGIC~I camera 
the PMTs have been divided into two categories: those which have a higher gain than 30000 when applying 850\,V and those with a lower gain.
The ones with a higher gain are attenuated in the PMT base reducing the signal amplitude by a factor of two. This had the effect that
the resulting gain and, therefore, also HV distributions are rather narrow. The quality of the HV flatfielding can be seen in Fig.~\ref{fig:charge}
and stays constant with
RMS/Mean at about 5-7\% RMS for months.
\\
%-------------------------------------------
\begin{figure}[htb]
\begin{center}
\includegraphics[width=1.0\linewidth]{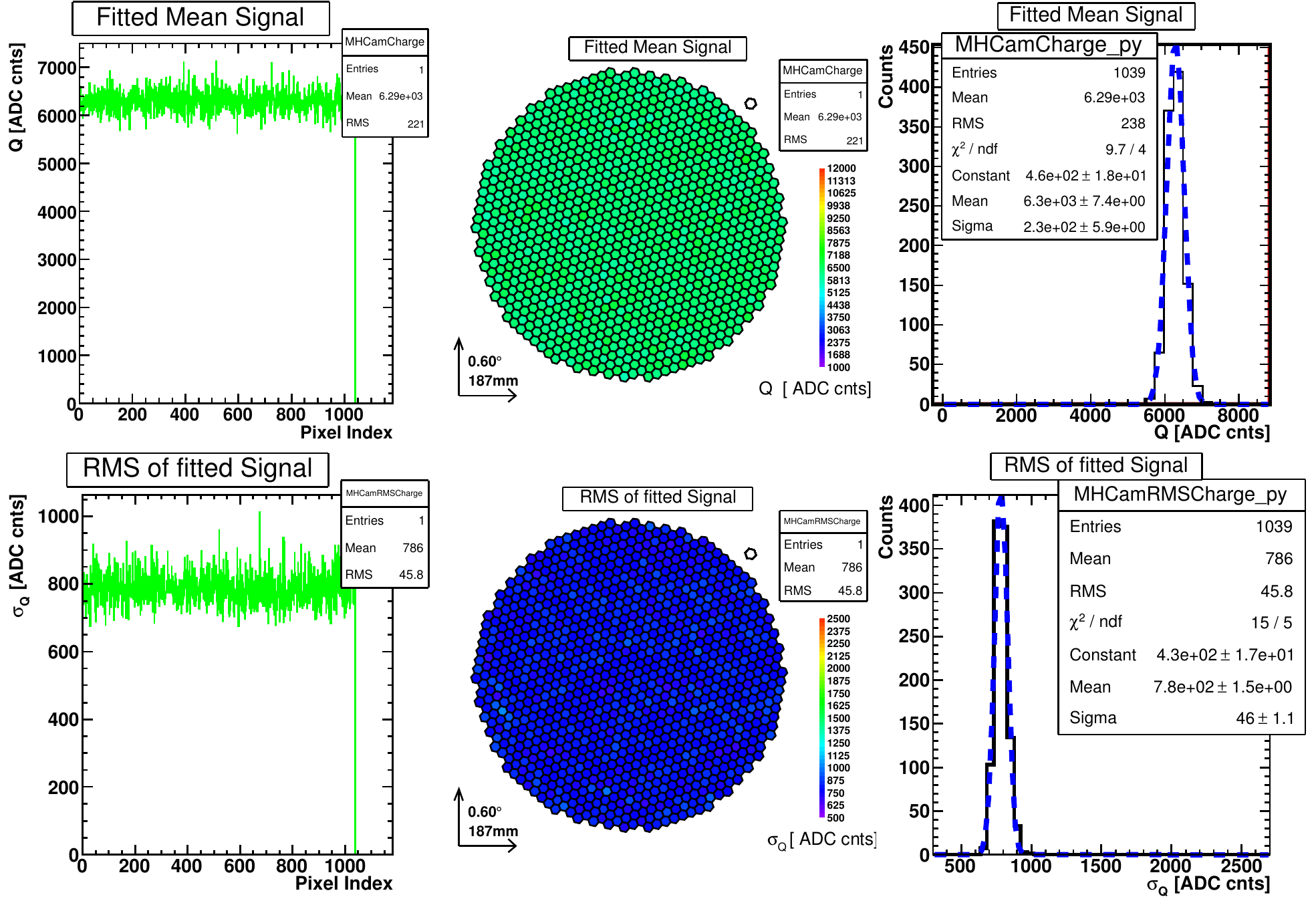}
\caption{
Rate scans taken changing discriminator thresholds to optimize the operating point of the MAGIC telescopes.
Red (blue) points are L1 3NN rate scans taken with MAGIC~I (MAGIC~II) telescopes and the lines are analytical fits to them.
The black points correspond the resulting stereoscopic rate of the system. The operating point has been chosen around 4.5\,phe
per channel.}
\label{fig:ratescan}
\end{center}
\end{figure}
%-------------------------------------------
{\bf L0 delays and L0 width adjustment.} Due to the different light propagation times of the signals from the focal plane of the camera
to the trigger system, the arrival times from different channels are slightly different (of the order of 2-3ns). This is mainly due 
to the differences in photo-electron propagation time in the PMTs when applying different HVs. 
An L0 delay is implemented on the receiver boards for each trigger channel in order to equalize arrival times of the signals
at the L1 trigger for photons arriving at the same time at the camera plane. A dedicated hardware and software has been built to
adjust the L0 delays automatically within several minutes. The delays are adjusted by injecting signals in each trigger pixel combination
(e.g. every 3NN combination) and varying the delays of the pixels to identify the parameter space producing valid macrocell trigger signals. 
Finally the delays are set to the best
values obtained from this scan. A minimum L0 signal width, which gave consistent results for all trigger channels, 
has been identified to be between 3ns to 5.5ns FWHM per pixel. 
In order to be on the safe side, all L0 trigger widths are then set to 5.5ns FWHM.
The resulting gate of the L1 trigger is about $(6\pm0.5)$\,ns.
\\
{\bf Discriminator threshold (DT) calibration}
The DTs of the trigger channels (L0 trigger) are adjusted such that the
sensitivity of the channels is flat in terms of photon density of Cherenkov
photons. This is achieved by means of a rate scan over the range of DTs for
each trigger channel when firing short (FWHM $<$2\,ns) calibration pulses with a given photon
density (e.g., equivalent to a mean of 100 photoelectrons (phe) per pixel of the PMT camera).
Then, for each trigger channel the required DT is determined, at which half of
the calibration pulses is accepted, and the other half is rejected.  In such a
way, the DTs are calibrated in terms of phe's.  We then scale the DTs linearly
to obtain a DT for a desired phe level.  As there are some small differences
between the analogue and digital signal chains, and the DT is applied to the
amplitude, whereas the HV flatfielding is done for the integrated signal, there
is some ~10\% spread of the resulting DTs (pixel-to-pixel).  
\\
{\bf Adjusting the operating point of the trigger}
Rates scans have been performed at clear nights at low zenith angles to determine what is
the shape of the trigger rate as a function of the DTs in phe. Mono rate scans as well as stereoscopic
rates scans have been performed and the performance has been shown to be stable for several nights.
An example of the rate scans is shown in Fig.~\ref{fig:ratescan}. One can see the steep slope of the rate at low DTs,
where the rate is dominated by the chance coincidence. At higher DTs, the trigger rate is dominated by the rate of the cosmic ray showers.
One can see that the coincidence trigger (stereo trigger) is strongly suppressing the chance coincidence triggers.
The operating point for MAGIC has been chosen to be around 4.5\,phe, resulting in a stereo rate of around 280 Hz,
whereas around 40Hz out of it is a chance coincidence trigger. 
\\
{\bf Individual pixel rate control (IPRC)}
The flatfielded
DTs may result in very different individual pixel rates (IPRs, measured on the
receiver boards) during the operation since a) the spectral sensitivity of PMTs
is different and b) the rates depend on the sky region the pixel is exposed to
(e.g. it may contain stars, which would increase NSB fluctuations and,
therefore, the IPR). As long as the IPRs are between 300kHz and 1.1MHz
(mean being around 800kHz), no action is taken. For IPRs outside of these
programmable limits, a IPR control software takes care of increasing or
decreasing of the DTs for the affected pixels in order not to spoil the
resulting L1 telescope rate.  

The performance of the upgraded system can be found in \cite{sitarek, icrc_performance}.
In addition to the upgrade as described in these proceedings, we are planning to install a
complementary trigger system, dubbed \emph{sumtrigger}, which has a lower energy threshold
and will operate parallel to the existing L1
digital trigger \cite{jezabel}. %, and will provide a substantial decrease in the energy
%threshold. The \emph{sumtrigger} can run in parallel with the Level-1 trigger and 
%its installation will not interfere with the
%standard operation and, therefore, no additional down time is expected. 

%-------------------------------------------

\section{Conclusions}

A major upgrade of the MAGIC telescopes took place in the last two years. The commissioning
of the upgraded system successfully finished in October 2012 and the telescopes
restarted regular operation. Besides an improvement of the sensitivity
(due to lower electronic noise in MAGIC~II, more homogeneous trigger response and
a larger trigger area of the MAGIC~I camera as well as due to the reduction of the dead-time), the main
goals of this upgrade were to secure stable performance and operation of the
telescopes for the next 5-7 years. % as well as to ease the maintenance of the system. 
The expectations concerning the sensitivity and the stability of the instrument were fulfilled \cite{icrc_performance}.
%the performance paper and first physics publications are under way.

\vspace*{0.5cm}
\footnotesize{{\bf Acknowledgment:}{
We would like to thank the Instituto de
Astrof\'{\i}sica de Canarias for the excellent working conditions at
the Observatorio del Roque de los Muchachos in La Palma. The
support of the German BMBF and MPG, the Italian INFN, the
Swiss National Fund SNF, and the Spanish MICINN is gratefully
acknowledged. This work was also supported by the CPAN
CSD2007-00042 and MultiDark CSD2009-00064 projects of the
Spanish Consolider-Ingenio 2010 programme, by grant 127740
of the Academy of Finland, by the DFG Cluster of Excellence
"Origin and Structure of the Universe", by the DFG Collaborative
Research Centers SFB823/C4 and SFB876/C3, and by the Polish
MNiSzW grant 745/N-HESS-MAGIC/2010/0.}}


\begin{thebibliography}{}

\bibitem{muxfadc} H. Bartko, F. Goebel, R. Mirzoyan, W. Pimpl, \& M. Teshima, 2005, 
NIM A, 548, 464.
\bibitem{bitossi} M. Bitossi, R. Paoletti and D. Tescaro, 2013, to appear in IEEE Transactions on Nuclear Sciences.
\bibitem{icrc_performance} J. Sitarek, M. Gaug, D. Mazin et al., NIMA in press, 2013, arXiv:1305.1007
\bibitem{sitarek} J. Sitarek, E. Carmona, P. Colin et al., 2013, Proc. of the 33rd ICRC, Rio de Janeiro, Id. 0074.
\bibitem{jezabel} J. Garcia Rodriguez et al., 2013, Proc. of the 33rd ICRC, Rio de Janeiro, Id. 666.

\end{thebibliography}
\end{document}